\documentclass[twocolumn,showpacs,preprintnumbers,amsmath,amssymb]{revtex4}

\usepackage{graphicx}
\usepackage{dcolumn}
\usepackage{bm}
\usepackage{amsmath}

\begin{document}
\author{G. Campagnano and Yu.V. Nazarov}
\affiliation{Kavli Institute of Nanoscience, Delft University of Technology, The Netherlands}

\title{$G_Q$-corrections  in Circuit Theory of Quantum Transport}

\date{\today}         

\begin{abstract} We develop a finite-element technique that allows one to
evaluate correction of the order of $G_Q$  to various transport
characteristics of arbitrary nanostructures. Common examples of such
corrections are weak localization effect on conductance and universal
conductance fluctuations. Our approach, however, is not restricted to
conductance only. It allows in the same manner to evaluate corrections
to noise characteristics, superconducting properties, strongly non-equilibrium 
transport and transmission distribution. To enable
such functionality, we consider Green
functions of arbitrary matrix structure. 
We derive a finite-element technique from Cooperon and
Diffuson ladders for these Green's functions.
The derivation is supplemented with application examples. 
Those include transitions between ensembles and Aharonov-Bohm effect.
\end{abstract}

\maketitle

\section{Introduction} The theoretical predictions of weak localization
\cite{WeakDiscovery} and universal conductance fluctuations
\cite{UCFDiscovery} along with experimental discoveries in this
direction \cite{GQExperimental,bergman} have laid a basis of modern
understanding of quantum transport --- transport in nanostructures ---
and have stimulated a considerable interest to the topic. Early studies
mostly concentrated on diffusive electron transport. Both effects arise
from quantum interference that is described in the language of slow
modes: Cooperons and Diffusons.\cite{Rama,Fukuyama} Each mode of this
kind brings a quantum (fluctuating) correction of the order of $G_Q
\equiv e^2/\pi\hbar$ to the classical Drude conductance $G$ of the
sample. This universal value sets an important division between
classical conductors ($G \gg G_Q$) where interference effects are small
and quantum ones ($G \simeq G_Q$) where the transport is essentially
quantum.

A complementary approach to $G_Q$-corrections comes
from Random Matrix Theory(RMT) of scattering \cite{Imry,BeenakkerReview}
This approach relates statistical properties of scattering matrix 
of a nanostructure to those of a certain ensemble of random matrices.
$G_Q$-corrections are understood in terms of fluctuations and rigidity of
spectral density of these matrices.
Although RMT approach can deal with diffusive systems, the most comprehensive
setup includes the so-called quantum cavity --- an element whose scattering matrix 
is presented by a 
completely random unitary matrix of a certain ensemble. The cavity can be seen 
as a region of space  where electron motion is sufficiently chaotic
(either ballistic or diffusive) 
and where electrons can get in and out through some constrictions \cite{cavity}.
The transport is determined by the propagation in the constrictions 
while
random unitary matrix representing the cavity is responsible for
"randomization" of the scattering. The RMT approach does not necessary 
concentrate on the total conductance. One can work with transmission distribution:
the averaged density of eigenvalues of transmission matrix squared.
This transmission distribution appears to be useful in a much broader
physical context: it determines not only conductance of nanostructures,
but also noise, full counting statistics of charge transfers, and properties of the 
same nanostructure with superconducting leads attached 
\cite{BeenakkerReview,NazarovOhm}. It is a modern paradigm of quantum transport
that an individual nanostructure is completely 
characterized by a set of transmission eigenvalues while transmission distribution
describes the averaged properties of random nanostructures of the same design.
This makes it relevant to study $G_Q$ corrections and fluctuations of
transmission distribution \cite{Carlo-corrections,Yuli-corrections}.
The density of transmission eigenvalues is of the order of $G/G_Q$, and
$G_Q$ corrections are of the order of $\delta G/G_Q \simeq 1$.

The microscopic Cooperon-Diffuson description is equivalent to
a proper RMT approach. This is best illustrated in the framework of
more general supersymmetric theory \cite{Efetov} that allows for
non-perturbative treatment of fluctuations in quantum scattering. 
The Cooperons and Diffusons in this theory are fluctuations of
supersymmetric field around the saddle point. 
For quantum cavity, only a single
mode of these fluctuations is relevant. 
The integration over these modes
reproduces RMT results \cite{Efetov,Zirnbauer,Iida}.

One can describe nanostructures in the framework of a simple
finite-element approach usually termed "circuit theory". The circuit
theory has originated from attempts to find simple solutions of Usadel
equations in superconducting heterostructures \cite{circuit-theory-old}.
However, it has been quickly understood that theories of the same
structure can be useful in much broader context: one can compute
transmission distribution \cite{NazarovOhm,ct-yuli}, noise and counting statistics
\cite{fcs-yuli}, investigate
spin effects \cite{spin} and non-equilibrium phenomena\cite{non-eq}.
In  circuit theory, a
nanostructure is presented in a language similar to that of electric
circuits: It consists of nodes, reservoirs and connectors. A node is in
fact a quantum cavity, a connector can be of very different types:
tunnel junction, ballistic contact and diffusive wire and is generally
characterized by a set of transmission coefficients.
In  circuit theory, each node is described by a matrix related to the
electron Green's function. In the limit $ G \gg G_Q$ the circuit theory
provides a set of algebraic equations that allow one to express 
the matrices in the nodes
in terms of fixed matrices in the reservoirs.

In this article, we present a technique to evaluate $G_Q$-corrections
for arbitrary nanostructures described by circuit theory. To allow for
various applications, we consider $G_Q$ corrections to multi-component
Green's functions of arbitrary matrix structure. We are able to present
$G_Q$ corrections in a form of a determinant made of derivatives of
Green's functions in the nodes with respect to self-energy parts. In terms
of a supersymmetric sigma-model, this corresponds to expansion of the
action up to quadratic terms. However, the formulation we present does
not contain any anti-commuting variables that complicate applications of
sigma-models. The determinant is just that of a finite matrix, 
this facilitates the computation of $G_Q$ corrections for
nanostructures of complicated design.

The structure of the article is as follows. To make it self-contained,
we start with a short outline of circuit theory of multi-component 
Green's functions
adjusted for the purposes
of further derivations. In Section \ref{sec:micro} we derive microscopic
expressions for $G_Q$ corrections and specify to finite-elements in
Section \ref{sec:answer}. Next Section \ref{sec:plugins} 
is devoted to description
of spin-orbit scattering and magnetic-field de-coherence that
are used to describe transitions between different RMT ensembles.
Since Aharonov-Bohm effect plays an important role in experimental
observation of $G_Q$ corrections, we explain how to incorporate it 
to our scheme in a short separate section \ref{sec:AB}.
We illustrate the technique with several examples 
(Section \ref{sec:examples})
concentrating on
a simplest $2\times2$ matrix structure that suits to calculate
$G_Q$ corrections to transmission distribution.
The examples also involve simplest circuits: One with a single node
and two arbitrary connectors, a chain of tunnel junctions, and two-node
four-junction circuit to demonstrate Aharonov-Bohm effect.
We summarize in Section \ref{sec:conclusions}.

\section{From Green's functions to finite elements}
\label{sec:circuit-theory}

In this article, we consider  Green's functions of arbitrary
additional matrix structure with $N_{ch}$ indices.
We do this for the sake of generality: This allows for 
description of super-conductivity, incorporating
Keldysh formalism and treating non-equilibrium and time-dependent
problems. This is also extremely convenient, since
the most relations in use do not depend on the "physical meaning"
of the structure. We use "hat" symbol 
for operators in coordinate space and "check" symbol for
matrices in additional indices. The Green's function
thus reads: $\hat G \equiv \check{G}(\bm{x},\bm{x}')$, 
$\bm{x}$ stands for (three-dimensional)
coordinate.
The general Green's function is defined as the solution of the following
equation 

\begin{equation} \left[ -\check \Sigma(\bm{x})-\hat H \right]\check
G(\bm{x},\bm{x}')=\delta(\bm{x}-\bm{x}'). \label{green}\end{equation} 

All physical quantities of interest can be in
principle calculated from
Green's functions. 
We address  here the quantum
transport of electrons in disordered media. 
In this case, one can work with common Hamiltonian 
\begin{equation}
 \hat H=\hat H_0 +u(\bm{x}), \,\,\,  \hat H_0=-\frac{1}{2m}\nabla^2_{\bm{x}}+U(\bm{x}).
\end{equation} 
Here  $U(\bm{x})$ describes the {\it design} of  nanostructure:
Potential "walls" that determine its shape, and form
ballistic quantum point contacts, potential barriers
in tunnel junctions, etc.
The potential $u(\bm{x})$ is random: it describes the random
impurity potential responsible for diffusive motion
of electrons, isotropization of electron distribution
function 
and, most importantly for this article, fluctuations
of transport properties of the nanostructure.
The physics at space scales exceeding the isotropization length
(which is the mean free path in the case of diffusive transport)
does not depend on a concrete model of randomness of
this potential.
Most convenient and widely used 
model assumes normal distribution of $u(\bm{x})$ 
characterized by the correlator 
$\langle u(\bm{x})u(\bm{x}') \rangle=w\delta(\bm{x}-\bm{x}')$.
It is important for us that both $\hat H_0$ and $u(\bm{x})$
are diagonal in check indices.

Most evident choice of self-energy matrix is
$\check\Sigma(\bm{x}) = -\epsilon$, $\epsilon$ being energy
parameter of the Green's function. $\check \Sigma$ is
more complicated in the theory of superconductivity.
We will find it
convenient, at least for derivation purposes,
to work with arbitrary $\check \Sigma(\bm{x})$. We can also consider
more general situation with non-local $\check \Sigma(\bm{x},\bm{x'})$.

Provided the conductance of the nanostructure is
sufficiently high ($G \gg G_Q$), one can disregard
quantum $G_Q$-corrections and work with semiclassical 
{\it averaged} Green's functions. Closed equations for those 
are obtained with the non-crossing approximation \cite{AGD}.
They include the impurity self-energy, 
\begin{eqnarray}
\left[
-\check \Sigma(\bm{x})-\check \Sigma_{imp}-\hat H_0 \right]\check
G(\epsilon;\bm{x},\bm{x}')=\delta(\bm{x}-\bm{x}'), \\ 
\check \Sigma_{imp}(\bm{x}) =w \check G(\bm{x},\bm{x})  
\end{eqnarray}
At space scales exceeding the mean free path, one can 
write closed equations for the Green's function in the coinciding points 
\cite{Usadel,Larkin}.
It is convenient to change notations introducing dimensionless  
$\check G(\bm{x}) \rightarrow  G(\bm{x},\bm{x})/i \pi \nu$, with  $\nu$ the density of states
at Fermi energy.
For purely diffusive transport, one obtains Usadel equation 
\begin{equation}\label{usadel}
\nabla \cdot \check{\bm{j}} - i \pi \nu [\check{\Sigma}(\bm{x}),\check{G}(\bm{x})]=0, \ \ \ \
\check{\bm{j}} = -\pi D(\bm{x}) \nu \check{G} \bm{\nabla} \check{G},
\end{equation}
$D$ being the diffusion coefficient.
The solutions of Usadel equations are defined only if one takes into account
boundary conditions at "infinity": Equilibrium Green's functions in the macroscopic 
leads adjacent to the nanostructure. It turns out that $\check G$ satisfies 
unity condition $\check G^2 = \check 1$. For situations where the transport
is not entirely diffusive, one would supplement the Usadel equation
with boundary conditions of various kinds (c.f. \cite{Kuprianov}).

An alternative way to proceed is to notice that the Usadel equation 
is almost a conservation law for the matrix current $\check{\bm{j}} $,
which allows for a finite element approach.
This conservation law is exact at space scale smaller than
the coherence length estimated as $\sqrt{D \Sigma}$. It does not relay
on assumption of diffusivity: It occurs because  the 
Hamiltonian $\hat H$ commutes with the "check" structure.
It is then natural  to proceed to the separation of the nanostructure into
regions where $\check G(\bm{x})$ can be assumed constant.
The next steps are the same as in traditional circuit theory of electric circuits,
which exploits conservation of electric current  and finite discretization
of the system onto regions of approximately constant voltage --- {\it nodes}.
Each node is connected by {\it connectors} to other nodes or reservoirs,
those representing macroscopic leads.

After the discretization of the nanostructure we can write a  Kirchhoff-like equation
\begin{eqnarray} \label{Kir}
 \sum_{c} \check I_c + \sum_\alpha \check{I}_{l,\alpha} =0,
\end{eqnarray}
where the indices $c$ and $\alpha$ label  the nodes and the
connectors of the nanostructure. 
The current through a connector $c$ which connects 
nodes $c1$ and $c2$ reads

\begin{equation}
\label{connector-curr}
\check{I_{c}} =\frac{1}{4}\sum_n \frac{T^c_n[\check G_{c1},\check G_{c2}]}
{1+T^c_n(\check G_{c1}\check G_{c2} +\check G_{c2}\check G_{c1}-2)/4}, 
\end{equation}
$\{T^c_n\}$ being the set of  transmission eigenvalues of the transmission matrix
squared relative to the connector. Eq. \ref{usadel} shows that 
the current $\check{\bm{j}}$ is not  fully conserved; to this aim we included
in Eq. \ref{Kir} a {\it leakage} current 
\begin{equation}
\check{I}_{l,\alpha} =- \frac{i\pi}{\delta_\alpha }[\check \Sigma_{\alpha}, \check{G}_{\alpha}], 
\ \ \ \ \ \ \delta_\alpha = \frac{1}{\nu V_\alpha} ,
\end{equation}
where $V_\alpha$ is the volume of the node; 
$\delta_\alpha$ can be easily recognized as the average 
level spacing in the node. 

It is opportune to notice that the conservation law (\ref{Kir}) can be 
obtained by requiring that the values  assumed by the matrix Green's function in the nodes 
are such to minimize the following the action
 \begin{eqnarray}
{\cal S}=\sum_c {\cal S}_c +\sum_\alpha {\cal S}_\alpha \label{action-ct}\\
{\cal S}_c = \frac{1}{2} \sum_n \mbox{Tr} \ln \left[ 1+\frac{T_n}{4}(\check G_{c1}\check G_{c2} +\check G_{c2}\check G_{c1} -2)\right]   \label{actionconnector} \\
{\cal S}_\alpha=-\frac{i\pi}{\delta_\alpha} \mbox{Tr} \check \Sigma_{\alpha}\check{G}_{\alpha}. \label{actionnode}
\end{eqnarray}
The minimization of the action must be carried out provided that the Green's functions
in each node satisfy the normalization condition $\check G_\alpha^2=1$; this implies that
the variation of the Green's function $\delta \check G_\alpha$ has to anti-commute with $\check G_\alpha$ 
itself. A possible way to satisfy this condition is to write the variation as 
\begin{equation}
\delta \check G_\alpha= \check G_\alpha  \delta \check v_\alpha- \delta \check v_\alpha \check G_\alpha,  
\end{equation}   
where no restriction is imposed on $\delta \check v_\alpha$.
The Kirchhoff equations are then re-written as
\begin{equation}
\frac{\delta \cal{S}}{\delta \check{v}_\alpha} =[\check{G},
\frac{\delta \cal{S}}{\delta \check{G}_\alpha}]=0.
\end{equation}
The same action can be used at microscopic level too,
even before averaging over $u(\bm{x})$.
To do this, we note once again that 
$\check\Sigma(\bm{x})$ form , at least formally, a set of parameters 
of our model. This can be straightforwardly extended to
non-local operators $\Sigma(\bm{x},\bm{x}')$.
To this end, we define the action in terms of the following variational formula
\begin{equation}
\delta {\cal S} = - \int d\bm{x}d\bm{x'}{\rm Tr} 
\left[ 
\check{G}(\bm{x}',\bm{x})\delta\check\Sigma(\bm{x},\bm{x}')\right],
\label{action-variation}
\end{equation}
which is traditional in Green's function applications\cite{AGD,Luttinger}.
With making use of formal operator traces, this action can be
written in terms of either $\hat \Sigma$ or  $\hat G$,
\begin{eqnarray}
{\cal S} = {\rm Tr} \ln \left( \hat \Sigma +\hat H\right);\label{Ssigma}\\
{\cal S} = {\rm Tr} \left[ -\ln \left( -\hat G \right) -
( \hat \Sigma +\hat H)\hat G \right]; \label{Sg}
\end{eqnarray}
apart from a constant. Eq. \ref{green} is reproduced
by varying either (\ref{Ssigma}) or (\ref{Sg}) over $\hat \Sigma$
or $\hat G$ respectively.
Since we are not planning to treat $G_Q$ corrections beyond perturbation
theory, we do not attempt to use the exponent of this action as
an integrand in some path integral representation 
of a result of exact averaging over $u(\bm{x})$. This has been
done in \cite{Efetov} for supersymmetric matrices $\check Q$
and in \cite{Larkin-great} for Keldysh-based two-energy matrices
$\check Q(\bm{x},\epsilon, \epsilon')$. We note, however, that
the action used in \cite{Efetov,Larkin-great} is equivalent 
to (\ref{action-ct}) if one substitutes $\check G = \check Q$. 

To this end, for averaged Green's functions the action in use is defined
simply as 
\begin{equation}
\delta {\cal S} = - \int d\bm{x}d\bm{x'}{\rm Tr} 
\left[ 
<\check{G}(\bm{x}',\bm{x})>\delta\check\Sigma(\bm{x},\bm{x}')\right].
\label{action-variation-averaged}
\end{equation}

\section{$G_Q$-corrections to multi-component Green's functions}
\label{sec:micro}
\begin{figure}

\includegraphics[scale=.5]{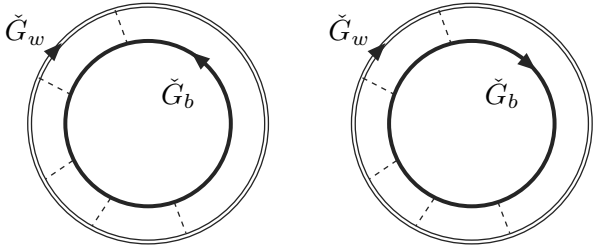}
\caption{ Wheel diagrams that determine $G_Q$-corrections to the action.
Single (double) line represents the Green's functions from the {\em black} ({\em white })  block while dashed lines represent the averaging over disorder. 
The  Diffuson (left) and Cooperon (right) wheel differ by 
mutual orientation of the lines.}
\label{fig:wheels}
\end{figure}

In this Section, we outline a microscopic approach to $G_Q$ corrections
suitable for multi-component Green's functions. The main idea
is the same as in \cite{Yuli-corrections,
Yuli-Annalen} where similar derivation
has been done for purely diffusive case
and for  a concrete $2 \times 2$
matrix structure.
It is known \cite{Rama,Fukuyama} that diffuson and cooperon modes
responsible for $G_Q$ corrections 
are presented as ladder diagrams made from averaged Green's functions.
In usual technique, such diagrams 
have  vertices corresponding
to two current operators in Kubo formula. 
One does not have to work with vertices: Instead, 
one considers Cooperon and Diffuson contributions to
the action that do not have any. 
The part of the action that presents $G_Q$ corrections
is given by wheel diagrams. These wheels
are made of either Cooperon or Diffuson 
ladder sections in a straightforward way (see Fig. \ref{fig:wheels}) 

We argue that the optimal way to present this part of the 
action is to {\it double} the existing "check" structure
and to consider Eq. \ref{green} in so-extended setup. Indeed,
suppose we would like to address the most general application 
of $G_Q$ corrections: Parametric correlations \cite{parametric}.
In this case, we start with two different "worlds" corresponding
to two different sets of parameters. In our case, all parameters can
be in principle incorporated into $\check \Sigma(\bm{x})$. So we 
have "white" and "black" sets, $\check \Sigma_{b}, \check \Sigma_{w}$,
those define corresponding non-averaged Green's functions
$\check G_{b}, \check G_w$. The averaging over the random potential $u({\bm x})$
provides correlations between the two "worlds" and gives the rungs of the
ladder diagrams that involve "white" and "black" Green's functions
(Fig. \ref{fig:wheels}). This is to obtain the Diffuson ladder wheel.
The Cooperon ladder  wheel is obtained by inverting the direction of the Green's
function in one of the "worlds".
This corresponds to using {\it transposed}
self-energies for this world, $\check\Sigma_w \rightarrow (\check\Sigma_w)^T$.
This guarantees that the corresponding Green's
functions are also transposed.
Fluctuations at the same values
of the parameters are naturally given
by diagrams where $\check\Sigma_b$ and $\check\Sigma_w$  either are
the same in both "worlds" (Diffusons), or mutually transposed
(Cooperons).

Finally, we note that the doubled structure is also useful for
evaluation of the weak localization correction. In this case, 
the last section of the Cooperon ladder is twisted before closing the
wheel. 

To proceed further, let us introduce the operators $\hat K$ 
presenting  a section of a corresponding ladder,
\begin{eqnarray}
\hat{K}_{\rm diff}^{ab,cd}(\bm{x},\bm{x'})= w(\bm{x}) 
\langle G_b^{ac}(\bm{x},\bm{x}')
\rangle \langle G_w^{db}(\bm{x'},\bm{x} )\rangle \\
\hat{K}_{\rm cooper}^{ab,cd}(\bm{x},\bm{x}')= w(\bm{x})
\langle G_b^{ac}(\bm{x},\bm{x}')\rangle
\langle G_w^{bd}(\bm{x},\bm{x}' )\rangle
\end{eqnarray}
where Latin letters represent "check" indices.
As we have noted, the "white" 
Green's function is transposed for the Cooperon. Those
are operators in the space spanned by the coordinates and the
 two check indices.

Summing up all diagrams, we find
the formal operator expressions for contributions to the action.
For fluctuations, we have
\begin{equation}
{\cal S}_{G_Q} = {\rm Tr}\left[ \ln ( 1 - \hat{K}_{\rm cooper})
 \right]+ {\rm Tr}\left[ \ln ( 1 - \hat{K}_{\rm diff}) \right]
\label{sgq}
\end{equation}
diffuson and cooperon contributions are
given by $\hat{K}_{diff}$, $\hat{K}_{cooper}$ respectively.
For weak localization correction, one has to account for the fact that the last
ladder section is twisted. We do this by introducing the permutation operator $\check{P}$,
which exchange "check" indices,
\begin{equation}
\hat{P}\, \hat{K}^{ab,cd}=\hat{K}^{ba,cd}.
\end{equation}
The contribution to the action becomes
\begin{equation}
{\cal S}_{wl} = \frac{1}{2}{\rm Tr}\left[
\hat{P} \ln (1 - \hat{K}_{\rm cooper})
\right].
\end{equation}
The factor one half is included in the last formula to take into account that "black" and "white" Green's functions, in the case of weak localization, are not anymore independent. 
We note that $\hat{K}$ for the Cooperon is symmetric with respect to index exchange,
so that $\hat{K}$ and $\hat{P}$ commute. The eigenfunctions and the eigenvalues of 
$K$ are therefore either symmetric ($K^{+}$)
or anti-symmetric ($K^{-}$) with respect to permutations.
We can rewrite the  last expression
as a sum over these eigenvalues 
\begin{equation}
{\cal S}_{wl} =\frac{1}{2} \sum_{n} \left[ \ln (1-K^{+}_n)- \ln (1-K^{-}_n)\right].
\label{s-wl}
\end{equation}

It is clear from the previous discussion that,
in order to calculate the $G_Q$ 
corrections, one has to evaluate
the eigenvalues of the ladder section $\check{K}$, both for
Cooperon and Diffuson.
We introduce now a method to compute this matrix easily.
The observation is that the ladders  under consideration are
not specific for $G_Q$-corrections: The same ladders determine a response
of {\it semiclassical} Green's functions upon variation 
of $\check \Sigma$ \cite{remark}.

\begin{figure}
\includegraphics[scale=.75]{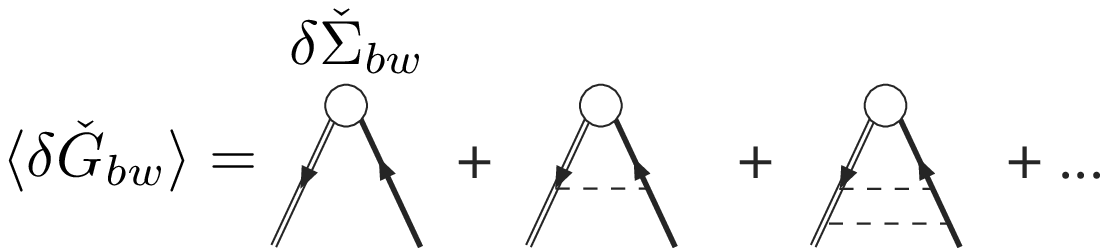}
\caption{The response of Green's function $\delta \check G_{bw}$ on self-energy $\check \Sigma_{bw}$ in semiclassical non-crossing approximation
is determined by the sum of ladder diagrams. This allows to express the wheel diagrams in terms of eigenvalues of response kernels. }
\label{fig:response}
\end{figure}

To see this, let us go back to non-averaged Green's functions.
We keep in mind that we have doubled the "check" space to include 
{\it white} and {\it black} sector. 
We add by hand a source term: The self-energy which mixes up 
"black" and "white" Green's functions,
$\delta\check{\Sigma}_{bw}(\bm{x})$. 
This source term will give rise to a 
correction to the Green's function in the same black-white sector.
In the first order, we have
\begin{multline}\label{response}
\delta\check{G}_{bw}(\bm{x},\bm{x}')=  \\
-\int d\bm{x}_1 d\bm{x}_2 \check{G}_{b}(\bm{x},\bm{x}_1)
\delta\check{\Sigma}_{bw}(\bm{x}_1,\bm{x}_2) \check{G}_{w}(\bm{x}_1,\bm{x}'),
\end{multline}
which is best illustrated by the diagram in Figure \ref{fig:response}. 
Next step is to include the effect 
of the random potential $u(\bm{x})$. We average Eq. \ref{response} 
limiting ourselves to the non-crossing approximation and obtaining
a set of ladder diagrams (Fig. \ref{fig:response}.)
By summing up all the contributions we obtain  
the correction---taken in coinciding points---to the Green's function
\begin{eqnarray}\label{av-response}
\langle \delta\check{G}_{bw}(\bm{x},\bm{x}) \rangle= 
\frac{1}{w(\bm{x})}\frac{\hat{K}_{\rm diff}}{1-\hat{K}_{\rm diff}}
\delta\check{\Sigma}_{bw}\, (\bm{x}).
\end{eqnarray}
Eq. (\ref{av-response}) is very valuable: it demonstrates
that the response of the 
Green's function to the source term $\delta\check{\Sigma}_{bw}$ 
is determined by the same ladder operator $\hat{K}_{\rm }$, 
which we need  to compute $G_Q$ corrections.

At the space scale of isotropization length
$\hat{K}\sim 1$. Usually one is interested in the contribution
arising from the larger space scale where Cooperon-Diffuson approximation
is valid. At this scale, the eigenvalues of $\hat K$ are either zero or very close
to one. To see this, we cite the results for the homogeneous case with  
$\check{\Sigma} ={\rm const}(\bm{x})$. A convenient basis in "check"
space is one where $\check\Sigma$ is diagonal, the eigenvalues being $\Sigma_n$.
The Green's function is diagonal in this basis as well, 
$G_n = s_a \equiv \rm{sign \ Im}(\Sigma_a)$. Owing to homogeneity,
the section operator is diagonal in the wave vector representation,
its eigenvalues being $K^{nm}(\bm{q})$. Direct calculation similar to
\cite{Fukuyama} gives $K^{nm}(\bm{q})=0$ if $s_n = s_m$. If $s_n \ne s_m$,
\begin{equation}
1-K^{ab}(\bm{q}) \approx \tau\left(i s_b(\Sigma_a -\Sigma_b) + D\bm{q}^2 \right) + ...
\end{equation}
for $\Sigma\tau, ql \ll 1$($\tau=2\pi w \nu$ is the isotropization time). This equation
makes the relation between our technique and the common technique for Cooperons
and Diffusons in homogeneous media. Usually, the self-energy $\Sigma$ has
equal number of eigenvalues with positive and negative imaginary parts.
In this case, at each $\bm{q}$, $\hat K$ has $N_{ch}^2/2$ zero 
and $N_{ch}^2/2$ non-zero eigenvalues.

Now we note that zero eigenvalues contribute neither to (\ref{sgq}) nor
to (\ref{av-response}). As to those close to $1$, we may replace $\hat
K$ by $1$ in the numerator of (\ref{av-response}). We also note that
$\delta \check G$ can be presented as the derivative of the action (c.f.
Eq. \ref{action-variation}). Therefore, we can write the $G_Q$
corrections due to the diffuson modes to the action in terms of a
determinant made of derivatives of the semiclassical action with respect
to 
$\check\Sigma_{bw},\check\Sigma_{wb
}$ \begin{equation} {\cal
S}_{G_Q,{\rm diff}}=-\ln \det{'}\left(-w(\bm{x}) \frac{\delta^2 {\cal
S}}{\delta\check{\Sigma}_{wb}\delta\check{\Sigma}_{bw}}\right).
\label{Smicroscopic} 
\end{equation}
The "prime" sign of the determinant signals that 
the zero eigenvalues shall be excluded: $det'$ is defined as the product
of all non-zero eigenvalues. Indeed, as we have seen, some variations
of self-energies do not change the Green's functions giving rise to zero eigenvalues. We also note
that the $G_Q$ corrections are not affected by the concrete form of
$w(\bm{x})$: Since the determinant of matrix product is a product
of their determinants, this matrix gives a constant contribution
to the action which does not affect any physical quantities.

\section{Method}   
\label{sec:answer}
Let us now adapt the microscopic relation (\ref{Smicroscopic})
to the finite-element approach outlined in the Section 
\ref{sec:circuit-theory}.

With all previous derivations, this step is easy. We just replace the
actual $\bm{x}$-dependent $\check{G}$ and $\check{\Sigma}$ by constants in
each node.
To get the action in these terms,
one integrates over the volume of each node so that
the formula (\ref{action-variation})
reads 
\begin{equation}
\delta {\cal S} = - \sum_{\alpha} \frac{i\pi}{ \delta_\alpha} {\rm Tr} 
\left[ 
\check{G}_{\alpha}\delta\check\Sigma_\alpha\right],
\label{action-variation-circuit}
\end{equation}
where the summation is over the nodes.
The discrete analog of the determinant relation (\ref{Smicroscopic})
is now 

\begin{multline} 
{\cal S}_{G_Q,{\rm diff}}  = -\ln \det{'}\left( -\frac{w_\alpha} {V_\alpha}
\frac{\delta^2 {\cal S}}{\delta\check{\Sigma}_{wb}\delta\check{\Sigma}_{bw}}\right) \\
-\ln \det{'}\left( \frac{\pi}{2 \tau_\alpha \delta_\alpha} 
\frac{\delta^2 {\cal S}}{\delta\check{\eta}_{wb}\delta\check{\eta}_{bw}}\right) 
 = {\rm const}-\ln \det{'}\left(  
\frac{\delta^2 {\cal S}}{\delta\check{\eta}_{wb}\delta\check{\eta}_{bw}}\right) 
\label{Scircuit}
\end{multline} 
where we have introduced dimensionless response matrix $\check \eta_\alpha \equiv i \pi \Sigma/\delta_\alpha$ and noticed that the matrix $\propto w$ brings a constant
contribution to the action. The response matrix is determined
from the solution of the Kirchhoff equations at vanishing source term $\check{\eta}_\alpha^{bw}$. It has $N_{nodes} \cdot N_{ch}^2/2$ non-zero eigenvalues and the same number of zero ones.
We observe that at $\Sigma_{w,b}=0$ the eigenvalues of this 
matrix do not depend on the volume of the nodes, they are determined by the transmission eigenvalues of the connectors only, and are of the order of
$G/G_Q$. Since rescaling of all conductances gives only an irrelevant constant contribution to the action, the $G_Q$ corrections depend only on the ratios
of conductances of the connectors: This manifests the universality of
these corrections.

The circuit theory action (\ref{action-ct})is given in terms of $\check{G}_\alpha$.
It is advantageous to present the answer for ${\cal S}_{G_Q}$ in terms of the expansion
coefficients of the action around the saddle-point: 
The solution of semiclassical circuit-theory equations. That is, to use $\delta^2 {\cal S}/{
\delta\check{G}_{wb}\delta\check{G}_{bw}}$ instead ${\delta^2 {\cal S}}/{\delta\check{\eta}
_{wb}\delta\check{\eta}_{bw}}$. 
If the latter matrix were invertible, we would make use of the fact that
$\delta^2 {\cal S}/{\delta\check{G}_{wb}\delta\check{G}_{bw}}=({\delta^2 {\cal S}}/{\delta\check{\eta}_{wb}\delta\check{\eta}_{bw}})^{-1}$. 
In fact, owing to the constrain $\check{G}^2=1$, there is a big number of zero eigenvalues
in the response matrix. So the task in hand is not completely trivial.

We proceed as follows.  
We expand the action by replacing each $\check{G}$ in each node by  
\begin{equation}
\check{G} = \check{G}_0 + \check{g}-\check{G}_0 \check{g}^2/2 + ...
\label{Gexpand}
\end{equation}
and collecting terms of the second order in $\check{g},\check{\eta}$. 
The form (\ref{Gexpand}) satisfies the constrain $\check{G}^2=1$ up to
second order terms provided $\check{g}\check{G}_0 +\check{G}_0\check{g} =0$.
Let us work in the $N_{ch}^2N_{node}$-dimensional 
space indexed with the "bar" index $\bar{a}$
composed of two "check" indices and one node index, $\bar{a} \equiv (a,b,\alpha)$. 
We present the result of the expansion as 
\begin{equation}
\delta {\cal S} = g^{wb}_{\bar{a}} M_{\bar{a}\bar{b}} g^{bw}_{\bar{b}} - 
\eta_{\bar{a}}g^{bw}_{\bar{a}}.   
\label{M-action}
\end{equation}
The variation of (\ref{M-action}) under the constrain 
$\check{g}\check{G}_0 +\check{G}_0\check{g} =0$ gives
 the response matrix
$\delta^2{\cal S}/\delta\eta_{\bar{a}} \delta\eta_{\bar{a}}$.
Next we consider matrix $\Pi_{\bar{a}\bar{b}}$
defined trough the following relation:
\begin{eqnarray}
\Pi_{\bar{a}\bar{b}} g^{bw}_{\bar{b}}\rightarrow \nonumber\\
\frac{1}{2}\left(\check{g} - \check{G}_0 \check{g} \check{G}_0\right)\rightarrow \nonumber\\
\frac{1}{2}\left(\check{g}^{bw} - \check{G}_b \check{g}^{bw} \check{G}_w \right),
\end{eqnarray}
the last equation makes white-black block separation explicit. We note
that $\Pi_{\bar{a}\bar{b}}$ is a {\it projector}: It separates "bar" space
on two subspaces where $\check{g}$ either commutes or anti-commutes with $\check{G}_0$,
and projects an arbitrary $\check{g}$ onto anti-commuting subspace.
Applying this projector to Eq. \ref{M-action}, we show that the projected
matrix $\Pi_{\bar{a}\bar{b}} M_{\bar{b}\bar{c}} \Pi_{\bar{c}\bar{b}}$ is
an inverse of the response matrix {\it within} the anti-commuting subspace. 
\begin{eqnarray}
{\cal S}_{G_Q} = \ln {\rm det}' \left(\Pi_{\bar{a}\bar{b}} M_{\bar{b}\bar{c}} \Pi_{\bar{c}\bar{b}}) \right) = \nonumber \\
\ln {\rm det} \left(\Pi_{\bar{a}\bar{b}} M_{\bar{b}\bar{c}} \Pi_{\bar{c}\bar{b}} +\delta_{\bar{a}\bar{b}}-\Pi_{\bar{a}\bar{b}} \right)
\label{in-M-and-P}
\end{eqnarray}
In the last equation we add the matrix $1-\hat{\Pi}$. 
This procedure replaces all zero eigenvalues with $1$, so one can evaluate a usual 
determinant.

We remind that as far as fluctuations are concerned,
there are two contributions of this kind coming from Diffuson
and Cooperon ladder respectively. The weak localization correction
involves permutation operator that sorts out eigenvalues involved
according to (\ref{s-wl}).
With this, the equations (\ref{in-M-and-P}) and (\ref{Scircuit})  give the $G_Q$ corrections
in an arbitrary circuit-theory setup in the most general form.
\section{Decoherence and Ensembles}
\label{sec:plugins}
Until now we have assumed that the Hamiltonian commutes with the "check"
structure and is invariant with respect to time reversal. 
This implies strict coherence of waves with different "check"
index which propagate in the disordered media described by this Hamiltonian.
Even small "check"-dependent perturbations of the symmetric Hamiltonian give
accumulating phase-shifts to these waves and may significantly change their interference patterns
at long distances. Due to its random nature, such phase-shifts 
can be regarded as decoherence although
this should not be confused with a real decoherence coming from interaction-driven
inelastic processes \cite{altrev2}.

In real experimental situations, two sources of such decoherence are usually of
importance: spin-orbit scattering and magnetic field.
Already early studies of $G_Q$ corrections \cite{WeakDiscovery,bergman} have revealed 
their significant dependence on these two factors in the regime where those
are too weak to affect the semiclassical transport. 
From the RMT point of view, these factors, upon increasing their strength,
provide transitions form orthogonal ensemble (symmetric Hamiltonian) to two different ensembles:
symplectic (spin-orbit interaction) and unitary (magnetic field) \cite{Efetov}.

\begin{figure}
\includegraphics[scale=.5]{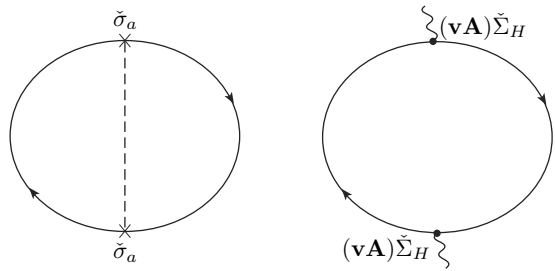}
\caption{Second-order diagrams in 
spin-orbit interaction (left) and magnetic field (right)
provide decoherence terms in the action. These terms describe transitions between pure RMT ensembles.}
\label{fig:plugins}
\end{figure}

We show in this Section how to incorporate spin-orbit scattering
and magnetic field into our scheme.
The most convenient way is to present them as perturbative
corrections to $\check{G}$-dependent action (Fig. \ref{fig:plugins}.) 

The spin-orbit scattering
enters the Hamiltonian in the form $\hat{H}_{so} = \check\sigma_{a} H^{a}(\bm{x},\bm{x'})$,
$H^{a}(\bm{x},\bm{x'}) = -H^{a}(\bm{x'},\bm{x})$, $\check\sigma_{a}$ representing
spin Pauli matrices in "check" space. In the second order in $H^{a}$ the averaging  gives (Fig. \ref{fig:plugins})
\begin{equation}
{\cal S}_{so} = \int d{\bm x} \frac{\pi \nu}{8 \tau_{so}(\bm{x})}{\rm Tr} \left[ \check{G}(\bm{x}) \check\sigma_{a} \check{G}(\bm{x}) \check\sigma_{a} \right]
\label{so-media}
\end{equation}
At the level of microscopic approach, the spin-orbit scattering takes
place anywhere in the nanostructure.
In the finite-element approach, it is advantageous to ascribe spin-orbit scattering
to nodes rather than to connectors. This is consistent with the main idea of our scheme:
Random phase-shifts take place in the nodes. 
The spin-orbit contribution in each node $\alpha$ is obtained by integrating (\ref{so-media})
over the node,
\begin{equation}
{\cal S}_{so} = \frac{\eta_{so}}{4} {\rm Tr}\left[ \check{G}_\alpha \check\sigma_{a} \check{G}_\alpha \check\sigma_{a} \right]
\label{decoherence-so}
\end{equation}
where $\eta_{so} \equiv \pi/(2\tau^{(so)}_\alpha \delta_\alpha)$.

The magnetic field is incorporated into the Hamiltonian through the modification of the derivative,
\[ i \nabla \rightarrow i \nabla- \frac{e}{c\hbar} \check{\Sigma}_H \bm{A} ,\]
where $\bm{A}$ is the vector potential and $\check{\Sigma}_H$ $(\check{\Sigma}^2_H = 1)$ 
describes the interaction of different "check" waves with the magnetic field.
In its simplest form, $\check{\Sigma}_H$ is the matrix in the white-black
structure introduced such that $\check{\Sigma}^b_H = 1$, $\check{\Sigma}^w_H = -1$
{\it provided} we describe a Cooperon. This is consistent with the requirement
that one of the Hamiltonians must be transposed to describe a Cooperon ladder. 
This is not the only plausible form of this matrix.
For instance, in non-equilibrium superconductivity $\check{\Sigma}_H$ involves 
electron-hole Nambu structure.

The magnetic field decoherence contribution can also be assigned to a node
and reads 
\begin{equation}
{\cal S}_H =\frac{\eta_{H}}{2} {\rm Tr}\left[ \check{G}_\alpha \check\Sigma_{H} \check{G}_\alpha \check\Sigma_{H} \right]
\label{decoherence-mf}
\end{equation}
where $\eta_H = \pi/(\tau_H \delta_\alpha)$ and  $\tau_H$ corresponds to 
Cooperon magnetic decoherence time in a common theory.
The latter is known  
to depend on the geometry of the node and its characteristics \cite{ABGreview}.
If the transport within the node is diffusive, $1/\tau_H = 4(e/\hbar)^2 D \langle\bf{A}^2\rangle$
where $\langle\rangle$ denote averaging over the volume of the node.
 The vector potential is taken in the gauge 
where it is orthogonal to the boundaries of the node. 
We have order of magnitude, $1/ \tau _H \delta \simeq (\Phi/\Phi_0)^2 (G_{node}/G_Q)$  
$\Phi$ being magnetic flux through the node, $\Phi_0 \equiv \pi \hbar/e$ being flux quantum,
and $G_{node}$ being a typical conductance of the node. The latter is limited by its Sharvin value
in the ballistic regime where the isotropization length is of the order of the node size.

The magnetic field produces not only random but also deterministic phase-shifts.
This gives rise to the Aharonov-Bohm effect discussed in the next Section.

To find the effect of decoherence terms (\ref{decoherence-so}) and (\ref{decoherence-mf}) 
on the eigenvalues forming the localization correction, 
we expand the action as done to obtain Eq. \ref{M-action}.
The decoherence contribution to $\hat M$ is diagonal in node index, 
and can be made diagonal in "bar" index by proper choice of the basis in "check" space.
For instance, if no external spin polarization is present in the structure,
spin-orbit contribution is diagonal in the basis made of singlets and triplets
in spin space. The simple realization of $\Sigma_H$ mentioned is automatically diagonal.
If in addition this diagonal contribution is the same in all nodes, both decoherence
effects just shift the eigenvalues of $\hat M$ corresponding to the symmetric Hamiltonian. 
This gives an extremely convenient model of decoherece effects.

The action for fluctuations is modified as follows:
\begin{eqnarray}
{\cal S}_{Diff} = \sum_{n} \ln(M_n) + 3 \ln(M_n + \eta_{so}) \\
{\cal S}_{Cooper} = \sum_{n} \ln(M_n+\eta_H) + 3 \ln(M_n + \eta_{so}+\eta_H)
\end{eqnarray}
where summation goes over non-degenerate eigenvalues of $\hat M$ and factor $3$ comes from three-fold degeneracy
of the triplet. 
To derive the modification for the week localization contribution, we note that
singlet and triplet are respectively antisymmetric and symmetric with respect to
permutation. Therefore, triplet extensions of symmetric and antisymmetric eigenvalues
are respectively symmetric and antisymmetric. The weak localization correction thus reads 
\begin{equation}
{\cal S}^{wl} = \frac{1}{2}\sum_n \ln\frac{M^-_n +\eta_H}{M^+_n +\eta_H}  
+\frac{3}{2} \ln \frac{M^+_n +\eta_{so}+\eta_H}{M^-_n +\eta_{so}+\eta_H}
\label{wl-with-shifts}
\end{equation}
$M^{+(-)}$ being (anti)symmetric eigenvalues of $\hat M$.

Since eigenvalues of $\hat M$ are of the order of $G/G_Q$, the decoherence effects
become important at $\eta_{so},\eta_H \simeq G/G_Q$, that is, when inverse decoherence
times match Thouless energy $E_{th}= (G/G_Q) \delta$ of the node, $1/\tau_{so},1/\tau_{H} \simeq E_{th}$.

\section{Ahronov-Bohm effect}
\label{sec:AB}

\begin{figure}
\includegraphics[scale=.6]{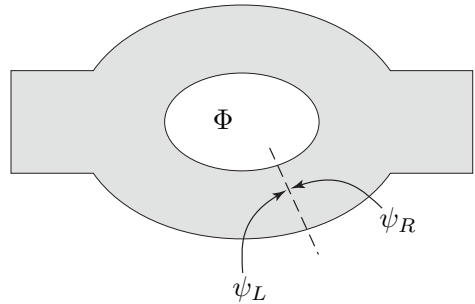}
\caption{ Ahronov-Bohm effect: the topological phase arising from the magnetic flux $\Phi$
is incorporated into the boundary conditions that relate the wave functions
on two sides of an imaginary cut. The cut can be made in any place.}\label{fig:ABring2}
\end{figure}

Aharonov-Bohm effect plays a crucial role in experimental observation and identification
of $G_Q$ corrections  (see e.g. (\cite{aron}).
Therefore it must be incorporated into our scheme, and in this Section we explicate how to do this.
This extends the results of Ref. \cite{stoof} where AB effect was considered in superconducting
circuit theory. In the following, we do not consider any orbital effects of the magnetic field but just 
the topological one. 

Let us suppose that the nanostructure presents a closed ring  threaded by a magnetic flux $\Phi$. 
As explained above, in presence of magnetic field the momentum operator has to be modified according  to
\[ i \nabla \rightarrow i \nabla- {e/\hbar} \check{\Sigma}_H \bm{A} ,\]
where $\bm{A}$ is the vector potential.  
Neglecting orbital effects,
one can get rid of the vector potential in the Schr\"odinger equation by a gauge transformation.
Let us have a ideal cut in the nanostructure that breaks the loop (Fig. \ref{fig:ABring2}.)
The topological effect of the flux can be incorporated into 
a boundary condition for the wave-function 
$\psi_{L,R}$ on two sides of the cut,
$\psi_L = \exp(i \phi_{AB}\Sigma_H)\psi_R$. 
The phase of the wave function therefore presents 
a discontinuity at the cut that is equal to $\pm \phi_{AB}, 
\phi_{AB}= \pi\Phi /\Phi_0$.
Since the transformation does not explicitly depend on $\bm{x}$, it can be immediately
extended to semiclassical  Green's functions. So that, those functions at two sides 
are related by
\begin{equation}
\label{gaugerotation}
\check{G}_L=\exp(i \phi_{AB} \check{\Sigma}_H)\check{G}_R\exp(-i \phi_{AB} \check{\Sigma}_H).
\end{equation}   

This solves the problem at microscopic level. 
Once the nanostructure has been discretized to finite elements, 
we note that the cut always occur {\it between} a connector and a node.
The most convenient 
way to deal with the gauge transformation (\ref{gaugerotation}) 
is to put it into the action of the corresponding connector.
To do this, we observe that the Green's function at the right end of the connector
is not $\check{G}_R$ of the node anymore: since the cut is crossed, it is eventually
$\check{G}_L$ given by (\ref{gaugerotation}). The connector 
action in the presence of
flux is therefore 
\begin{eqnarray}
S_c = \frac{1}{2} \sum_n \mbox{Tr} \ln \left[ 
1+\frac{T_n}{4}(\check G_{c1} \check G_{c2}(\phi_{AB}) + \right. \nonumber \\
\left. + \check G_{c2}(\phi_{AB}) \check G_{c1}
-2)\right]  
\end{eqnarray} 
where 
\begin{equation}
\check{G}(\phi_{AB})=\exp(i \phi_{AB} \check{\Sigma}_H)\check{G}\exp(-i \phi_{AB} \check{\Sigma}_H).
\end{equation}   
One checks that the variation of the so-modified action reproduces the
Kirchhoff laws for matrix current given 
in \cite{stoof}. Owing to global gauge invariance, it does not matter to
which connector and to which end of the connector the Aharonov-Bohm phase is ascribed.
If the are more loops in the nanostructure, more connector
actions have to be modified in such a way.

\section{Examples}
In the previous sections, 
we operated with general matrix "check" structure to keep the discussion as general as possible.
In this Section, we will give a set of examples to illustrate concrete applications of the technique developed.
For the sake of simplicity, we choose the simplest matrix structure that gives a sensible
circuit theory. 
We consider $2 \times 2$ matrix Green's functions
whose values in two terminals can be parametrized by a single parameter $\phi$,
\begin{equation}
\check{G}(\phi)=\left(
\begin{array}{cc}
0 &  e^{-i \phi} \\  
e^{i \phi} & 0
\end{array}
\right).
\label{two-by-two}
\end{equation}

The use of these matrix structures is that they give access to a fundamental quantity in quantum
transport: Transmission distribution of transmission eigenvalues of two-terminal nanostructure. We have to explain 
this relation before going to concrete examples.
The averaged transmission distribution is defined as 
\begin{equation}
\rho(T)=\sum_n \langle  \delta(T-T_n) \rangle
\end{equation}
where the sum is done over all the transport channels and 
the average is, in principle, to be intended over
an ensemble of nanostructures of the same design. 
For $G \gg G_Q$ the transmission eigenvalues are dense
in the interval $[0,1]$ and self-averaging takes place.

Let us take a connector and set Green's functions at its 
ends to $\check{G}(\phi_1),\check{G}(\phi_2)$. By virtue of
(\ref{actionconnector}) the connector action reads
\begin{equation}
{\cal S}(\phi_1 - \phi_2)=\sum_n \ln \left[ 1-T_n\sin^2 \frac{\phi_1 - \phi_2}{2}  \right].
\end{equation}

The trick is to regard the whole nanostructure as a single complex
connector between left and right reservoir, and set the Green's functions
in the reservoirs to $\check{G}(0),\check{G}(\phi)$. The total action (\ref{action-ct})
becomes now the connector action of the whole nanostructure and defines the transmission 
distribution in question,
\begin{equation}
{\cal S}(\phi)= \int dT \rho(T) \ln \left[ 1-T\sin^2 \frac{\phi}{2}  \right].
\end{equation}

If one computes the $\phi$-dependence of ${\cal S}$, the transmission distribution 
can be {\em extracted} from its analytic continuation on complex $\phi$ (\cite{NazarovOhm})
\begin{equation}
\label{distribution-extraction}
\rho(T)=-\frac{1}{\pi T \sqrt{1-T}}
{\rm Re}
\left[ \frac{\partial {\cal S}}{\partial \phi} (\pi +2 i \cosh^{-1}  \frac{1}{ \sqrt{T} } - 0^+)\right].
\end{equation}

Circuit theory of the Section \ref{sec:circuit-theory} gives the answer in the 
limit $G \gg G_Q$. The weak localization contribution ${\cal S}_{wl}$ gives
$G_Q$ correction to the transmission distribution. The fluctuation contribution ${\cal S}_{G_Q} = {\cal S}_{diff}
+{\cal S}_{cooper}$
that depends on two parameters $\phi_{w,b}$, gives {\it correlations} of transmission distributions
(c.f. \cite{Yuli-corrections}) 
\begin{eqnarray}\label{action-correlator}
{\cal S}_{G_Q}(\phi_b, \phi_w) = \int dT dT' \ll \rho(T) \rho(T')\gg \nonumber \\
\ln \left[ 1-T\sin^2 \frac{\phi_b}{2}  \right]
\ln \left[ 1-T'\sin^2 \frac{\phi_w}{2}  \right].
\end{eqnarray}
A simple application of the above formulas
are $G_Q$ corrections to the conductance. Those
are given by the derivatives of corresponding actions
at $\phi_{b,w} =0$,
\begin{eqnarray}
\frac{\delta G_{wl}}{G_Q} = -2\left.
\frac{\partial^2 {\cal S}_{wl}(\phi)}{\partial \phi^2} \right\vert_{\phi=0} \label{wl-conductance}\\
\frac{\ll G_{b}G_{w}\gg}{G^2_Q} = 4\left. 
\frac{\partial^4 {\cal S}_{G_Q}(\phi_b,\phi_w)}
{\partial \phi_b^2\partial \phi_w^2 }\right\vert_{\phi^b,\phi^w=0}  \label{fluct-conductance}
\end{eqnarray}

\label{sec:examples} \subsection{Junction Chain} 
\begin{figure}
\includegraphics[scale=.6]{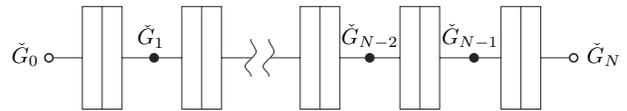}
\caption{ The chain of tunnels junctions of the same conductance $G_T$.
In our finite-element approach, there are two Cooperon and two Diffuson eigenvalues per node. In the limit of $N\rightarrow \infty$  the chain is shown to be identical 
to a continous quasi-one-dimensional diffusive wire.}
\label{fig:junctionchain}
\end{figure}

The first example is a chain of tunnel junctions.
We will study $G_Q$ corrections for a chain of
$N$ identical junctions that connects  two reservoirs (Fig.
\ref{fig:junctionchain}.)
 The
connector action for a tunnel junction assumes a very simple form:
${\cal S}_c = (G_T/4G_Q) {\rm Tr} \left(\check{G}_{1c}
\check{G}_{2c}\right)$, $G_T$ being conductance of the tunnel junction.
Several tunnel junctions in series, however, provide a good approximate
for diffusive wire. Therefore, in the limit $N \to
\infty$ we can compare $G_Q$ corrections with the known results
\cite{Carlo-corrections,Yuli-corrections} for corrections to
transmission distribution of a one-dimensional diffusive conductor.

We
set Green's functions in the reservoirs on the left and on the right to
$\check{G}(0),\check{G}(\phi)$ (c.f. Eq.\ref{two-by-two}) respectively.
The semiclassical action for the system reads
\begin{equation}\label{taction}
{\cal S}=\frac{G_T}{4 G_Q}
\sum_{k=0}^{N-1} {\rm Tr} 
\check{G}_k\check{G}_{k+1}- 
\frac{i \pi}{\delta_S} \sum_{k=1}^{N-1} {\rm Tr\check{\Sigma}_k \check{G}_k} ,
\end{equation} 
Here $k= 1, \dots, N-1$ labels the nodes, 
while $k=0$ and $k=N$ identify left and right reservoir respectively.
All the nodes are assumed to be identical with the same $\delta_S$.

Since all the junctions are identical, the semiclassical 
solution is easy to find: the "phase" $\phi$ drops by the same
amount at each junction, and the solution reads 
$\check{G}_k = \check{G}(k\phi/N)$, provided
$\check\Sigma_k \propto \check{1}$ within each $2\times2$ 
block. 
This gives the optimal value of the action,
\begin{equation}
{\cal S} =\frac{N G_T}{2G_Q}\cos\left(\frac{\phi}{N}\right)
\end{equation}
from which one can 
evaluate the semiclassical transmission distribution 
by using relation (\ref{distribution-extraction}).
In the limit $N \to \infty$ ${\cal S} = -({G_D}/{4 G_Q})\phi^2$,
$G_D \equiv G_T/N$ being the conductance of the whole chain.
This gives well-known transmission distribution for
diffusive conductor, $\rho(T) = (G_D/2G_Q)/T\sqrt{1-T}$ \cite{NazarovOhm}.

As explained above, to calculate  $G_Q$ corrections 
we augment the "check" dimension of the  
Green's functions by introducing the "black" and "white" structure. 
Consequently, the parameter $\phi$ gets a "color" 
index $b$ or $w$. The semiclassical solution for resulting $4\times4$
matrix is non-zero in $bb$ and $ww$ blocks,
\begin{equation}
\label{saddle-point-chain}
\check{G}^0_k=\left(
\begin{array}{cc}
\check{G}(k\phi_b/N) &  0 \\  
0 & \check{G}(k\phi_w/N)
\end{array}
\right).
\end{equation}

Now we shall derive the matrix $\hat M$ eigenvalues
of which determine $G_Q$ corrections. It is advantageous
to use a parametrization of the deviations 
 from the semiclassical solution,$\check{g}$, which
automatically satisfy $\check{g} \check{G} +\check{G} \check{g}=0$
in each node. To this end, we rewrite the action
(\ref{taction}) in a special basis: That one where 
$\check{G}^0_k$ are diagonal in each node,
\begin{equation}
\check{G}^0_k=
\left(
\begin{array}{cccc}
1 &0 & 0&0 \\
0 & -1& 0&0 \\
0 &0 & 1&0 \\
0 & 0& 0&-1 
\end{array}
\right).
\end{equation}
Then the deviation of the form 
\begin{equation}
\check g_k=\left(
\begin{array}{cccc}
0 &0 &0&g^{bw}_{k,p} \\
0 & 0& g^{bw}_{k,m}&0 \\
0 &g^{wb}_{k,p}& 0&0 \\
g^{wb}_{k,m} & 0& 0&0 
\end{array}
\right)
\end{equation}
satisfies the condition mentioned. 

In this basis (see Appendix \ref{AppA} for details)
the action reads 
\begin{equation}\label{rotaction}
{\cal S}=\frac{G_T}{2 G_Q}\sum_{k=0}^{N-1} {\mbox Tr}
 \check G_k \check L \check G_{k+1} \check L^{-1}
 -\frac{i \pi}{\delta_S}  \sum_{k=1}^{N-1} 
  {\mbox Tr} \check \Sigma_k \check G_k, 
\end{equation}
where $bb$ $(ww)$ block of $\check{L}$ is given by
\begin{equation}
L^{bb(ww)}=\left(
\begin{array}{cc}
\cos (\phi^{b(w)}/N) & i \sin (\phi^{b(w)}/N) \\
i \sin (\phi^{b(w)}/N) & \cos(\phi^{b(w)}/N) 
\end{array}
\right).
\end{equation}
We expand the Green's matrices according to Eq. (\ref{Gexpand}),
write the quadratic form 
in terms of $\check{g}$ and diagonalize it 
(see Appendix \ref{AppA})
to find the following set of eigenvalues, ($l = 1 \dots N-1$)
\[
\frac{4G_Q}{G_T}M^{\pm}_l(\phi^b,\phi^w)=
2\cos \frac{\phi^b}{2N}
\cos \frac{\phi^w}{2N}
\cos\frac{\pi l}{N}-\cos \frac{\phi^b}{N}
\]
\begin{equation}
-\cos \frac{\phi^w}{N}  
\mp \sqrt{4\sin^2 \frac{\phi^b}{2N}\sin^2 \frac{\phi^w}{2N}
\cos^2\frac{\pi l}{N}+\epsilon^2},
\end{equation}
$\epsilon \equiv 2\pi G_Q(\Sigma^b-\Sigma^w)/G_T i\delta_S$
measures the difference of Green's function energy parameter
in $bb$ and $ww$ blocks in units of a single-node Thouless energy.
To obtain eigenvalues that determine the weak localization contribution, 
we set $\phi^w=-\phi^b=\phi, \epsilon = 0$. This yields
\begin{equation}
\label{wlteigenal1} 
M^{+}_{wl,l}(\phi)=\frac{G_T}{4 G_Q}
\cos( \frac{\phi}{N})\left[ \cos \frac{\pi l}{N}-1\right]
\end{equation}
\begin{equation}
\label{wlteigenal2} 
M^{-}_{wl,l}(\phi)=\frac{G_T}{4 G_Q}
\left[ \cos \frac{\pi l}{N} -\cos \frac{\phi}{N}\right]
\end{equation}

Let us discuss the weak localization correction first.
If we neglect decoherence factors, 
we can sum up over $l$ to find a compact analytical expression
\begin{equation}
{\cal S}_{wl} = \frac{(N-2)}{2}\ln(\cos\frac{\phi}{N}) +\frac{1}{2}\ln
\left(\frac{\sin\frac{2\phi}{N}}{\sin \phi}
\right)
\label{chain-wl}
\end{equation}
In the limit $N \to \infty$ this reproduces 
the known correction 
for one-dimensional diffusive wire \cite{Yuli-corrections},
\begin{equation}
{\cal S}_{wl} = \frac{1}{2}\ln
\left(\frac{\phi}{\sin \phi}
\right)
\label{1d-correction}
\end{equation}
It is interesting to note that the weak localization 
correction is absent for $N=2$.
We will see below that 
this is a general property of a single-node
tunnel-junction system.
It was observed in \cite{Yuli-corrections} that a part of 
weak localization correction in diffusive conductors is universal:
It depends neither on the shape nor on the dimensionality of the conductor. 
The universal part is concentrated near transmissions close to one
and is given by
\begin{equation}
\delta\rho_{wl}(T) = -\frac{1}{4}\delta(T-1),
\label{wl-universal}
\end{equation}
while the non-universal part is a smooth function of $T$.
The relation (\ref{chain-wl}) possesses this property at
any $N$, since the universal part comes from
the divergency in (\ref{chain-wl}) at $\phi =\pi$ where the eigenvalue 
$M^-_{wl,1}$
goes to zero. 
Our approach proves that this correction is universal for a large
class of the nanostructures, not limited to diffusive ones,
for any nanostructure where transmission eigenvalues approach $1$.
This is guaranteed  by the  
logarithmic form of the action.
If $M^-_{wl,1} \propto (\pi -\phi)$ at $\phi \to \pi$,
the correction is given by (\ref{wl-universal}) irrespective of 
the proportionality coefficient.

Expanding (\ref{chain-wl}) at $\phi \to 0$ we find the
correction to the conductance of the tunnel junction chain,
\begin{equation}
\frac{\delta{G}_{wl}}{G_Q} = -\frac{1}{3} \frac{(N-1)(N-2)}{N^2}
\end{equation}
This is written for orthogonal ensemble, a well-known 
factor $(1-2/\beta)$ defines the correction for other pure ensembles. 
The effect of spin and magnetic decoherence can be taken into account 
by shifting the eigenvalues 
(\ref{wlteigenal1} ,\ref{wlteigenal2}) 
according to Eq. \ref{wl-with-shifts}
since the decoherence factors in each node are the same, 
$\zeta_{H,so} \equiv \eta_{H,so}(4G_Q/G_T)$.

\begin{figure}
\includegraphics[scale=.7]{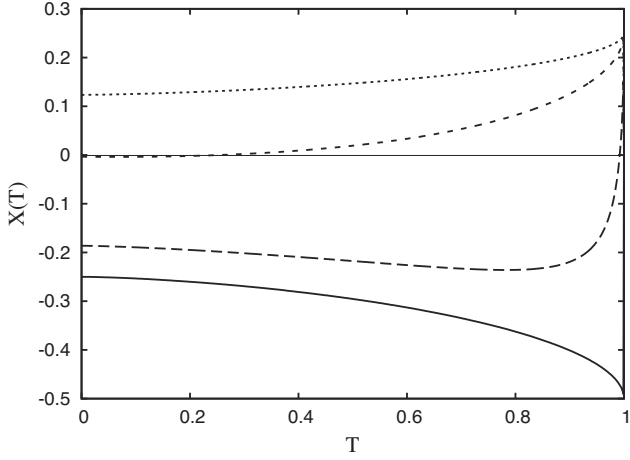}
\caption{
Weak localization 
correction to the transmission distribution 
of a system of four identical junctions at different values of spin-orbit 
parameter $\zeta_{SO}$. We plot here cumulate correction 
$X(T) \equiv \int_T^1 dT' T'\delta\rho(T')$. $X(1)$ represents the universal singular part 
of the correction (c.f. Eq. \ref{wl-universal}) while $X(0)$ gives correction to the conductance. The lowest
curve corresponds to strictly zero $\zeta_{SO}$ and therefore represents  pure
orthogonal ensemble. Its negative value at $T=1$
is partially compensated by positive non-universal contribution coming from
$T\simeq 1$ so that the resulting correction to conductance $\delta G_{wl}/G_Q = X(0) \approx 0.2$.
The two higher curves correspond to relatively small values
of $\zeta_{SO}$, $0.05$ and $0.4$. While they  are close to the orthogonal ensemble result
 at $T \simeq 1$, their behavior at $ T \approx 1$ is quite
different: the universal correction is that of symplectic ensemble
and is of positive sign. The highest curve corresponding to
$\zeta_{SO}=10$ is close the cumulate correction 
of pure symplectic ensemble, $X_{sym}(T) = - X_{ort}(T)/2$. 
}
\label{fig:weaksochain}
\end{figure}

The ${\cal S}_{wl}$ is 
still given by analytical although 
lengthy expression (See Appendix A). 
The correction to the transmission distribution corresponding
to this expression is plotted in Fig. \ref{fig:weaksochain} for different
strengths of spin-orbit coupling to illustrate the transition
between the orthogonal and simplectic ensembles.
The correction to the conductance is given by
\begin{equation}
\frac{\delta G_{wl}}{G_Q}=3F(N,\zeta_H+\zeta_{so})-F(N,\zeta_H)
\end{equation}
where we define an auxiliary function $F(x,N)$ 
\begin{multline*}
F(x,N)=
-\frac{(N-1)}{N^2}
-\frac{1}{N^2}\frac{(2+x)^2}{x(4+x)}
\\
-\frac{1}{N\sqrt{x(4+x)}}
\frac{2^{2N}+\left(2+x+\sqrt{x(4+x)}\right)^{2N}}{2^{2N}-\left(2+x
+\sqrt{x(4+x)}\right)^{2N}}
\end{multline*}

Let us discuss the parametric correlations. Without decoherence 
factors and at the same energy 
($\epsilon =0$) 
one can still sum up over the modes
to obtain an analytical expression,
\begin{multline}
{\cal S}_{Diff} = {\cal S}_{Coop} = (N-1)
\ln\left(\cos\frac{\phi_b}{N}+\cos\frac{\phi_w}{N}\right)
+  \\
+\sum_{\pm} \ln\left(\frac{\sin\frac{\phi_b \pm \phi_w}{2}}{\sin\frac{\phi_b \pm \phi_w}{2N}}\right)
\end{multline}
The fluctuation of conductance obtained with (\ref{fluct-conductance})
reads
\begin{equation}
\frac{\langle \langle \delta G^2 \rangle \rangle}{G^2_Q} = \frac{2}{15} 
\frac{N^4 +15
N -16}{N^4}
\end{equation}
and converges to the known expression for quasi-one-dimensional 
diffusive conductor at $N \to \infty$. We notice that this
convergence is rather quick, the fluctuation at $N=5$ differs
from asymptotic value by $10\%$ only. We see thus that 
the diffusive wire, that in principle contains an infinite number
of Cooperon and Diffuson modes, can be, with sufficient accuracy,
described by the finite-element technique even at low number of elements.

\begin{figure}
\includegraphics[scale=.7]{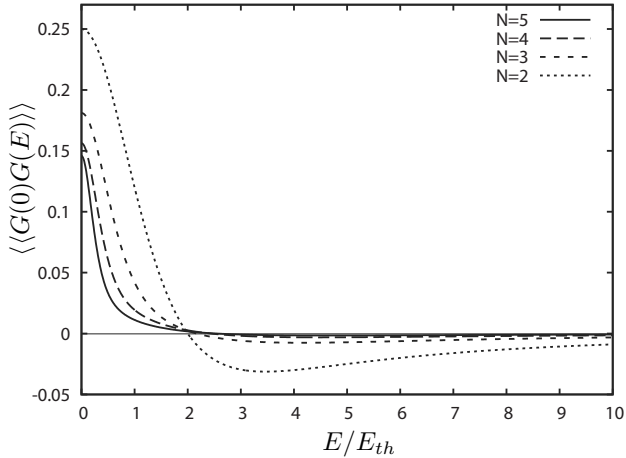}
\caption{ Energy dependence of
the correlator of conductance fluctuations $\ll G(0) G(E)\gg$ for
chains with different number of junctions $N$. The energy difference
is normalized to Thouless energy of the whole chain, $E_{th} \equiv  \delta_s G_T/2\pi G_Q N^2$.
Note the fast convergence of the correlator to that of diffusive wire
for big $N$ and negative correlations at large $E$.}
\label{fig:fluct-chain}
\end{figure}

Another point to discuss concerns the correlations
of transmission eigenvalues $T_n$, those can be
obtained by analytic calculation of (\ref{action-correlator}).
It is instructive to 
concentrate on the relatively small eigenvalue separations
those are much smaller than $1$ but still exceed an average
spacing $\simeq G_Q/G$ between the eigenvalues,
$G_Q/G \ll |T-T'|\ll 1$. We observe that 
the correlation in this case is determined by
the divergency of ${\cal S}$ at $\phi_b -\phi_w \to \pm 2\pi$.
Indeed, $M^{-}_{1}$ approaches $0$ in this limit.
This again suggest the universality of these correlations. 
Indeed, as shown in \cite{Yuli-corrections} for diffusive conductors,
the correlations in this parameter range are determined
by universal Wigner-Dyson statistics and reduce to 
\begin{equation} 
\label{small-correlation}
\langle \langle  \rho(T) \rho(T') \rangle \rangle =-\frac{2}{\pi^2\beta} {\rm Re}\frac{1}{(T-T'+i0^+)^2}.
\end{equation}
Since the conductance fluctuations are contributed 
by correlations of $T_n$ at scale $\sim 1$ as well,
they are not universal. We plot in Fig. \ref{fig:fluct-chain}
the correlator of conductance fluctuations as function of
energy difference at several $N$.
 
\subsection{AB ring}
\begin{figure}[h]
\includegraphics[scale=.3]{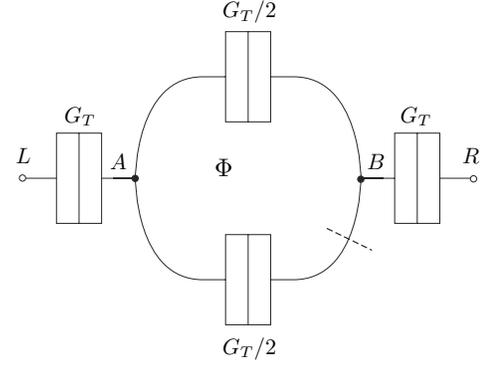}
\caption{A simplest circuit to exemplify AB effect consists of two nodes and
for (tunnel) connectors. AB phase modifies the Green's function on the right side 
of the lowest connector with respect to $\check G_B$. The cut is given by dotted line.}
\label{fig:AB-simple}
\end{figure}

In this subsection 
we exemplify evaluation of $AB$ effect within our scheme.
We concentrate on the simple circuit presented in Fig. \ref{fig:AB-simple}. 
It contains
four tunnel junctions and two nodes labeled 
$A$ and $B$. The conductances of the junctions are chosen
to re-use the results of the previous section for a chain of
three tunnel junctions: The solution of semiclassical circuit
theory equations is given by (\ref{saddle-point-chain}) for $N=3$.
The action reads 
\begin{eqnarray} 
S=\frac{G_T}{4G_Q}{\mbox Tr} 
\left\{  \check{G}_L\check{G}_A + 
\frac{1}{2} \check{G}_A\check{G}_B \right. \nonumber \\
\left.+\frac{1}{2} \check{G}_A\check{G}_B(\phi_{AB}) 
+ \check{G}_B \check{G}_R \right\} 
-i\frac {\pi}{\delta} \sum_{i=A,B} {\mbox Tr} \check{\Sigma}_i \check{G}_i. 
\label{ahronovaction} 
\end{eqnarray}  
where $\check{G}_{L,R}$ are Green's functions in the reservoirs and 
$\check G_B(\phi_{AB})$ is modified according to Eq. \ref{gaugerotation}.
To study correlation of conductance fluctuations, we consider 
different Green's functions for white and black block,
and subjected to different fluxes $\phi_{AB}^b \ne \phi_{AB}^w$. 
For weak localization correction, we set $\phi_{AB}^b = -\phi_{AB}^w =\phi_{AB}$

To calculate the matrix $\hat M$
we use again the basis 
where $G^{(0)}_i$ are diagonal 
and the parametrization 
for $\check g$ introduced in the previous subsection. 
It reads 
\[M= \left(
\begin{array}{cc}
M_d & M_{od} \\
M^*_{od} & M_d
\end{array}
\right)
\] 
where $2\times 2$ blocks $M_{d,od}$ are given
\begin{eqnarray*}
\frac{G_T M_d}{4 G_Q}=\left(
\begin{array}{cc}
-\cos\varphi^b-\cos\varphi^w+\epsilon  &  0 \\
0 & -\cos\varphi^b-\cos\varphi^w-\epsilon
\end{array}
\right)
\\
\frac{G_T M_{od}}{4 G_Q}=\frac{1+e^{i (\phi_{AB}^b-\phi_{AB}^w)}}{2} 
\left(
\begin{array}{cc}
\cos \frac{\varphi^b}{2}\cos \frac{\varphi^w}{2} &  \sin \frac{\varphi^b}{2}\sin \frac{\varphi^w}{2} \\
\sin \frac{\varphi^b}{2}\sin \frac{\varphi^w}{2}  &   \cos \frac{\varphi^b}{2}\cos \frac{\varphi^w}{2} 
\end{array}
\right)
\end{eqnarray*}
The parameter
$\epsilon$ which characterizes 
the energy difference between  
black and white Green's 
function is defined as in the previous subsection. 
At $\epsilon=0$ we obtain explicit 
expression for the Diffuson eigenvalues 
(the Cooperon ones are  obtained  by   
$\phi^w \rightarrow -\phi^w$ and $\phi_{AB}^w \rightarrow -\phi_{AB}^w$),

\begin{eqnarray}
\frac{4 G_Q}{G_T}M_{1,2}^+=-\cos\frac{\phi_b}{3}-\cos\frac{\phi_w}{3} \nonumber\\
\pm \cos\left(\frac{\phi^b_{AB}-\phi^w_{AB}}{2}\right)
\cos\left(\frac{\phi_b-\phi_w}{6}\right),
\\
\frac{4 G_Q}{G_T}M_{1,2}^-=-\cos\frac{\phi_b}{3}
-\cos\frac{\phi_w}{3} \nonumber\\
\pm \cos\left(\frac{\phi^b_{AB}-\phi^w_{AB}}{2}\right)
\cos\left(\frac{\phi_b+\phi_w}{6}\right).
\end{eqnarray} 

The weak localization correction to the action reads
\begin{equation}
{\cal S}_{wl}(\phi)=\frac{1}{2}\ln\left(\frac{ \cos^2(\phi/3)(4-\cos^2(\phi_{AB}))}{4 \cos^2(\phi/3)-\cos^2(\phi_{AB})}    \right),
\end{equation}
from which we calculate the correction 
to conductance as function of the flux
\begin{equation}\label{ab-weak}
\frac{\delta G_{wl}}{G_Q}=-\frac{4\cos^2(\phi_{AB})}{9(7-\cos(2\phi_{AB}))}.
\end{equation}
We see that the weak localization correction cancels 
at half-integer flux $\phi_{AB}=\pi$. This is because 
the junctions forming the loop are taken to be identical.
The flux dependence exhibits higher harmonics indicating
semiclassical orbits that encircle the flux more than once.

For the correlator of conductance fluctuations we obtain 
\begin{equation}
\frac{\langle \delta G^2 \rangle}{G^2_Q}=
\sum_{\pm}\frac{259-4\cos(\phi^w_{AB}\pm\phi^b_{AB}) +
\cos2(\phi^w_{AB}\pm\phi^b_{AB})}{81(\cos(\phi^w_{AB}\pm\phi^b_{AB})-7)^2}
\end{equation}
where plus and minus signs indicate Cooperon and Diffuson 
contributions respectively. The higher harmonics are present
as well.
\subsection{Two connectors and one node}

\begin{figure}
\includegraphics[scale=.7]{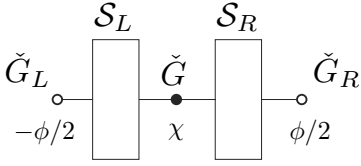}
\caption{The simplest possible circuit comprises one node, 
two connectors($L$ and $R$), and two reservoirs where Green's functions $\check G_{L,R}$ are fixed.}
\label{fig:two-conn}
\end{figure} 

Probably the simplest system to be considered by circuit theory methods
consists of a single node and two connectors (Fig. \ref{fig:two-conn}). 
Since in this case there are only $N_{ch}$ eigenvalues,
one can straightforwardly 
elaborate on complicated arbitrary connectors. For  this setup 
we are still able to find an analytical expression for Cooperon and 
Diffusion eigenvalues. This allows us to get an expression for the
weak localization correction to the conductance which was vanishing
in the case of two tunnel junctions.
Each connector is in principle characterized by the
distribution of transmission coefficients $\{T^R_n\}$, $\{T^L_n\}$,
or, equivalently, by the functional form of the connector 
action given by Eq. (\ref{actionconnector}), ${\cal S}_{L},{\cal S}_{R} $ .
The action for the whole system reads
\begin{equation}\label{2conn-action}
{\cal S}={\cal S}_{L}(\check G_L,\check G)+{\cal S}_{R}
(\check G,\check G_R)-i \frac{\pi}{\delta_s}  {\rm Tr} \check \Sigma \check G.
\end{equation} 
$\check G$ being the Green's function of the node. 
We employ $2\times2$ matrices parametrized by (\ref{two-by-two}), and set
the Green's functions in the left and right reservoir to 
$\check G(-\phi/2)$ and $\check G(\phi/2)$ respectively. 
The saddle-point value of $\check G$ is given by the phase $\chi$ 
and for a general choice of
${\cal S}_L$ and ${\cal S}_R$ does depend on 
$\phi$, $\chi \equiv \chi(\phi)$. The total action in the saddle point is therefore ${\cal S}(\phi) ={\cal S}_L(\chi+\phi/2) +{\cal S}_R(\chi-\phi/2)$. We expand the Green's function according to 
Eq. \ref{Gexpand}.
The second order correction to the action in this case reads
\begin{multline}
{\cal S}^{(2)} =-i \frac{\pi}{\delta_s}  \mbox{Tr}\, \check \Sigma \,\check g  \\
 -\frac{1}{2}  \sum_n \mbox{Tr} \Bigg\{  \frac{T^L_n}{4[1+\frac{T^L_n}{4}( \{ \check G^0, \check G_L \} - 2)]}  \check g^2 \check G^0  \check G_L  \\
  + \left[\frac{T^{L}_n}{4[1+\frac{T^L_n}{4}( \{ \check G^0, \check G_L \} - 2)]}\right]^2  \ 
(\check g  \check G_L \check g \,\check G_L + \check g^2)   + (L \leftrightarrow R) \Bigg\}. 
\end{multline}
Acting like in the previous subsections, we
find two following Diffuson eigenvalues (the Cooperon ones are
obtained  by the substitution  
$\phi^w \rightarrow -\phi^w$).
\begin{widetext}
\begin{multline}
M^\pm(\phi^b,\phi^w)=\sum_{i=b,w}I(\phi^i) \left[\cot[\chi(\phi^i)-\frac{\phi^i}{2}]-\cot[\chi(\phi^i)+\frac{\phi^i}{2}]  \right]  \\
-\frac{1-\cos[\chi(\phi^b)+\frac{\phi^b}{2}\mp\chi(\phi^w)\mp\frac{\phi^w}{2}]}{2}
\sum_{i=b,w}\left\{\frac{1}{\sin^2[\chi(\phi^i)+\phi^i/2]} \left[\frac{I'(\phi^i)}{\chi'(\phi^i)+1/2}-I(\phi^i)\cot[\chi(\phi^i)+\frac{\phi^i}{2}]  \right]
\right\}  \\
+\frac{1-\cos[\chi(\phi^b)-\frac{\phi^b}{2}\mp\chi(\phi^w) \pm \frac{\phi^w}{2}]}{2}\sum_{i=b,w}
\left\{\frac{1}{\sin^2[\chi(\phi^i)-\phi^i/2]} \left[\frac{I'(\phi^i)}{\chi'(\phi^i)-1/2}-I(\phi^i)\cot[\chi(\phi^i)-\frac{\phi^i}{2}]  \right]
\right\}. \label{two-connect-eigenvalues}
\end{multline} 
\end{widetext}
Here we introduce $I(\phi) \equiv \partial {\cal S}/{\partial \phi} $
to characterize the derivative of the total semiclassical action.
We see that $M^-$ approaches zero in the limit $\phi_b,\phi_w \to \pm \pi, \mp \pi$ provided $I(\phi)$ stays finite. As discussed, this divergency guarantees the universality of the correlations of transmission eigenvalues.

Below we  specify to three different cases.

\subsubsection{Symmetric setup}
If we set ${\cal S}_R={\cal S}_L$, 
  $\chi(\phi^{b(w)})$ is zero regardless the concrete form ${\cal S}_L$. 
The total action therefore reads ${\cal S}(\phi)=2 {\cal S}_L(\phi/2)$.
 The eigenvalues (\ref{two-connect-eigenvalues}) take a simpler form. 
To compute the weak localization correction to the action we set
 $\phi^b=-\phi^w=\phi$ to find
\begin{equation} \label{symmetric-wl-action}
{\mathcal S}_{wl}=\frac{1}{2}\ln\left(\frac{I'(\phi)}{I(\phi)}\tan\frac{\phi}{2}\right)+{\rm const}.
\end{equation}
For tunnel junctions, $I(\phi)\propto \sin (\phi/2)$ and the 
correction disappears.

Expanding Eq. \ref{symmetric-wl-action} in Taylor series near $\phi \rightarrow 0$ 
we reproduce the well-known result of \cite{Iida} for weak localization correction to conductance. To comply with the notations used there, we characterize connectors
with $t_p=\sum_n T_n^p$. 
\begin{equation}
\frac{\delta G_{wl}}{G_Q}=-\frac{ t_2}{4 t_1},
\end{equation} 
Similar expansion of Diffusion and Cooperon eigenvalues reproduces 
the result of \cite{Iida} for conductance fluctuations,
\begin{equation}
\frac{\langle (\delta G)^2 \rangle}{G_Q}=\frac{3t_2^2+2t_1^2-2t_1(t_2+t_3)}{8 t_1^2}. 
\end{equation}  

\subsubsection{Diffusive connectors}
It is instructive for understanding circuit theory of $G_Q$ corrections
to specify the relation (\ref{symmetric-wl-action}) 
to diffusive connectors.
Since in this case $I(\phi) \propto \phi$, we obtain
\begin{equation}
{\cal S}_{wl,node} = \frac{1}{2} \ln (\tan(\phi/2)/\phi)
\end{equation}
Two connectors, single node situation can be easily realized
in a quasi-one-dimensional wire with inhomogeneous resistivity distribution
along the wire. A low-resistivity region would make a node if bounded
by two shorter resistive regions that would make the connectors.
On the other hand  it has been proven in \cite{Yuli-Annalen}
 that the weak localization correction in inhomogeneous
 wires does not depend on the resistivity distribution.
 Therefore, it has to be universally given by (\ref{1d-correction}),
 ${\cal S}_{wl,1d} = (1/2) \ln (\phi/\sin \phi) \ne {\cal S}_{wl,node}$. 
How to understand this apparent discrepancy?

This illustrates a very general point: $G_Q$ corrections may be accumulated
at various space scales ranging from mean free path to sample size.
The experimental observation of the corrections relies on the ability to separate the
contributions coming from different scales, e.g. by changing magnetic field \cite{bergman}.
With our approach, we evaluate the part coming from interference
at the scale of the node. The part coming from interference
at shorter scale associated with the connectors is assumed to be included
into transmission distribution of these connectors.

For our particular setup, this extra contribution comes from
two identical connectors. Since only half of the phase $\phi$ drops at 
each connector, the contribution equals $2 {\cal S}_{wl,1d}(\phi/2)$.
Summing up both contributions, we obtain 
$$
{\cal S}_{wl,node} + 2 {\cal S}_{wl,1d}(\phi/2)={\cal S}_{wl,tot} ={\cal S}_{wl,1d}(\phi)  
$$
That is, the weak localization correction in this case remains universal 
provided the contribution of the node is augmented by contributions of two
connectors.

\subsubsection{Non-ideal Quantum Point Contact}
The transmission distribution of an ideal multi-mode Quantum Point 
Contact (QPC) with conductance $G_B \gg G_Q$ is very degenerate since all
$T_n=1$ or $0$. This degeneracy is lifted if the QPC is adjacent
to a disordered region, even if the scattering in this region is weak. 
This can be modeled as a connector with
conductance $G_D \gg G_B$ in series.

The weak localization correction to the conductance has been calculated
already a while ago (\cite{Beenakker-QPC}). In the relevant limit, it is
parametrically small in comparison with $G_Q$, $\delta G_{wl} =-G_Q (G_B/G_D)$.
Usual way to verify the applicability of semiclassical approach to
quantum transport is to compare the conductance of a nanostructure
with the weak localization correction to it. For generic nanostructure,
this gives $G \gg G_Q$. However, for our particular example $\delta G_{wl} \ll G_B$ even for a few-channel QPC where $G_B \simeq G_Q$. So the question is:
Is semiclassical approach really valid at $G_B \simeq G_Q$?

To answer this question, we compute the weak localization correction to the transmission distribution.
Since system is not symmetric, we make  use of the full expression 
(\ref{two-connect-eigenvalues}). In the limit of $G_D \gg G_B$, the
relevant values of $\phi$ are close to $\pi$. We stress it by shifting the phase, $\mu=\pi-\phi, |\mu| \ll 1$. 

The circuit theory analysis in semiclassical limit 
gives \cite{campagnano}
\[ I(\mu)=G_D(-\frac{\mu}{2}+\sqrt{\frac{\mu^2}{4}+R_c}) \hspace{1cm}  
\chi=\frac{\pi}{2}+\sqrt{\frac{\mu^2}{4}+R_c},\]
where $R_c\equiv 4 G_B/G_D \ll 1$. 
This gives to the following distribution
of reflection coefficients (c.f. Eq. \ref{distribution-extraction})
\[
\rho(R) = \frac{G_D}{2 \pi G_Q} \Theta(R_c -R)
\sqrt{\frac{R_c}{R} -1 }. 
\]

We use the above relations with (\ref{two-connect-eigenvalues}) to 
find the Cooperon 
eigenvalues,
\begin{eqnarray}
M^+(\mu) &= &-G_D\left(-\frac{1}{2}+\frac{\mu/4}{\sqrt{\frac{\mu^2}{4}+R_c}} \right)\frac{\frac{\mu^2}{4}+R_c}{R_c/2}, \\
M^-(\mu) &= &G_D\left(-\frac{\mu}{2}+\sqrt{\frac{\mu^2}{4}+R_c}\right)\frac{\mu}{2R_c}.
\end{eqnarray}
This yields the  weak localization correction to the current,
\[
I_{wl}(\mu)=\frac{2R_c}{\mu(\mu^2+4R_c)}.
\]
The resulting correction to the transmission distribution consists
of two delta-functional peaks of opposite sign, those are located
at the edges of the semiclassical distribution,
\begin{equation}
\delta \rho_{wl}(R)=\frac{1}{4}\left[\delta(R-R_c)-\delta(R)\right].
\label{qpc-correction}
\end{equation} 
To estimate the conditions of applicability, we smooth the correction
at the scale of $R_c$. This gives $|\delta \rho|/\rho \simeq G_Q/G_B$
and the semiclassical approach does not work at $G_B \simeq G_Q$. This agrees
with RMT arguments given in \cite{campagnano}. The correction to the conductance
calculated with (\ref{qpc-correction}) agrees with the value cited above and indeed is anomalously small.

\section{Conclusions}
\label{sec:conclusions}
We  present a finite-element method to evaluate quantum corrections --- 
typically of the  order of $G_Q$ --- to  transport characteristics of arbitrary 
nanostructures. This includes universal
conductance fluctuation and weak localization.
We work with  matrix Green's
functions of arbitrary structure to treat a wider class
of problems that includes superconductivity, full counting statistics and  non-equlibrium transport. At microscopic level, the corrections are expressed in terms of Diffuson
and Cooperon modes of continuous Green's functions.
We employ a variational method based on an action to formulate a consistent finite-element approach. 

We illustrate the method with a set of simple and physically interesting examples. All examples
are based on $2 \times 2$ matrices, this suffices to calculate the
transmission distribution of two-terminal structures. We show
how a chain of tunnel junctions approaches the diffusive wire
upon increasing the number of junctions and study transitions between ideal RMT 
ensembles in the chain. We consider the simplest finite-element system
that exhibits Ahronov-Bohm effect. We obtain general results for a single-node system  with two arbitrary connectors and check their consistence
with the well-known results for quantum cavity. This allows us to improve understanding of quantum interference in inhomogeneous diffusive wires and non-ideal quantum point-contacts.

\acknowledgments
We appreciate useful discussions with Ya. M. Blanter, P.W. Brouwer and
M.R. Zirnbauer. This work was supported by the Netherlands Foundation for Fundamental
Research on Matter (FOM.)

\appendix
\section{Junction chain}
\label{AppA}
In this  Appendix we illustrate how to find the eigenvalues of the matrix $M$ defined in section \ref{sec:answer} for the chain of tunnel junctions introduced. We consider the action in the rotated basis presented in text
\begin{equation}
{\cal S}=\frac{G_T}{4 G_Q}\sum_{i=0}^{N} {\mbox Tr} \check G_i \check L \check G_{i+1} \check L^{-1}-\frac{i \pi}{\delta_s}  \sum_{i=1}^{N-1}  {\mbox Tr} \check \Sigma_i \check G_i. 
\end{equation}
We expand the Green's matrices according to Eq. (\ref{Gexpand})

The second order correction to the action reads (the first order is zero because we perform the expansion around the stationary point)
\begin{widetext}
\begin{equation}
{ \cal S}^{(2)}=\frac{G_T}{4 G_Q}  \sum_{i=0}^{N} {\rm Tr}\left\{   \check g_i \check L \check g_{i+1}\check L^{-1}  -\frac{1}{2}\check g_i \check L\check G^{0}_{i+1} \check L^{-1}-\frac{1}{2} \check G^{0}_i \check L\check  g_{i+1}\check L^{-1}  \right\}
-\frac{i \pi}{\delta_s}  \sum_{i=1}^{N-1}  {\rm Tr}  \left\{ \delta \check \Sigma_i \check g_i -\frac{1}{2} \check \Sigma_i\check g^2_i\check G_i^{0} \right\}           
\end{equation}
\end{widetext}

Taking the variation respect to $g^{wb}_{p,k}$ and $g^{wb}_{m,k}$ leads to the eigenvalue equations the matrix $M$. 
\begin{widetext}
\begin{multline}\label{eigen1}
\sin \frac{\phi^b}{2N}\sin \frac{\phi^w}{2N} (g^{wb}_{p,k+1}+g^{wb}_{p,k-1})+\cos \frac{\phi^b}{2N}\cos \frac{\phi^w}{2N} (g^{wb}_{m,k+1}+g^{wb}_{m,k-1})-(\cos \frac{\phi^b}{N}+\cos \frac{\phi^w}{N})g^{wb}_{m,k} \\
-\frac{1}{2 \xi}(\Sigma^b-\Sigma^w)g^{wb}_{m,k}-\frac{4 G_Q}{G_T} M g^{wb}_{m,k}=0,
\end{multline}

\begin{multline}\label{eigen2}
\sin \frac{\phi^b}{2N}\sin \frac{\phi^w}{2N} (g^{wb}_{m,k+1}+g^{wb}_{m,k-1})+\cos \frac{\phi^b}{2N}\cos \frac{\phi^w}{2N} (g^{wb}_{p,k+1}+g^{wb}_{p,k-1})-(\cos \frac{\phi^b}{N}+\cos \frac{\phi^w}{N})g^{wb}_{p,k} \\
+\frac{1}{2 \xi}(\Sigma^b-\Sigma^w)g^{wb}_{p,k}-\frac{4 G_Q}{G_T} Mg^{wb}_{p,k}=0,
\end{multline}
\end{widetext}

where $\xi=i  \delta_s \, G_T / (4 \pi G_Q )$. To solve the system of coupled equations it is convenient to write
\begin{equation}\label{sol}
 g^{wb}_{p,k}=g^>_{p} e^{i q k} +g^<_{p} e^{-i q k},
\end{equation}
and similar equation for the $m$-component. The coefficient $q$ is to be determined from the boundary conditions 
\[
g^{wb}_{p,0}=0,
\] 
\[
g^{wb}_{p,N}=0.
\] 
From the boundary conditions we have $g^>_{k}=-g^<_{k}$ and $q=\pi l/N$ with $l=1,..,N-1$. We substitute the expressions \ref{sol} in the equations for the eigenvalues and get
\[
\frac{4G_Q}{G_T}M^{\pm}_l(\phi^b,\phi^w)=2\cos \frac{\phi^b}{2N}\cos \frac{\phi^w}{2N}\cos(\frac{\pi l}{N})-\cos \frac{\phi^b}{N}
\]
\begin{equation}
-\cos  \frac{\phi^w}{N}  \mp \sqrt{4\sin^2 \frac{\phi^b}{2N}\sin^2 \frac{\phi^w}{2}\cos^2(\frac{\pi l}{N})+\epsilon^2}
\end{equation}
The quantity $\epsilon$ being $(\Sigma^b-\Sigma^w)/2 \xi$. 

Here we report the correction to the action when decoherence and spin orbit are present as generalization of Eq. \ref{chain-wl} , let us define the following functions
 \begin{multline*}
s_1(\phi,x,N)=\frac{1}{2}\left(2+\frac{x}{\cos(\phi/N)}\right)  \\
+\sqrt{\frac{1}{4}\left(2+\frac{x}{\cos(\phi/N)}\right)^2-1} 
 \end{multline*}
 \begin{multline*}
s_2(\phi,x,N)=\frac{1}{2}\left(x+2\cos(\phi/N)\right)
\\
+\sqrt{\frac{1}{4}\left(x+2 \cos(\phi/N) \right)^2-1} 
 \end{multline*}
 and the function 
 \begin{multline*}
 A(\phi,x,N)=  \\
 \left[\frac{s_1^2-s_1^{-2}}{2-s_1^2-s_1^{-2}} +N\frac{-s_1^{2N}+s_1^{-2N}}{2-s_1^{2N}-s_1^{-2N}}\right]
 \frac{\partial}{\partial \phi} \ln \left(s_1(\phi,x,N)\right) \nonumber \\
 - \left[\frac{s_2^2-s_2^{-2}}{2-s_2^2-s_2^{-2}} +N\frac{-s_2^{2N}+s_2^{-2N}}{2-s_2^{2N}-s_2^{-2N}}\right]\frac{\partial}{\partial \phi} \ln \left(s_2(\phi,x,N)\right)
\end{multline*}
We can finally express the correction to the action as
\begin{equation}
\frac{\partial}{\partial \phi} {\cal S}_{wl}(\phi) = A(\phi,\zeta_H,N)-3 A(\phi,\zeta_H+\zeta_{so},N)
\end{equation}


\begin{thebibliography}{1}

\bibitem{WeakDiscovery}
L. P. Gor'kov, A. I. Larkin and D. E. Khmel'nitskii,
Pis'ma Zh. Eksp. Teor. Fiz. 30, 248 (1979) [JETP Lett. 30, 228 (1979)];
E. Abrahams, P. W. Anderson, D. C. Licciardello, and
T. V. Ramakrishnan, Phys. Rev. Lett. {\bf 42}, 673
(1979).

\bibitem{UCFDiscovery} B. L. Altshuler, Pis'ma Zh. Eksp. Teor. Fiz. {\bf 41},
530 (1985) [JETP Lett. {\bf 41}, 648 (1985)];
P. A. Lee and A. D. Stone, Phys. Rev. Lett. {\bf 35}, 1622 (1985);
B. L. Altshuler and D. E. Khmelnitskii, JETP Lett. {\bf 42}, 559 (1985).

\bibitem{GQExperimental} 
S. Washburn and R. A. Webb, Rep. Prog. Phys. {\bf 55}, 1311 (1992). 

\bibitem{bergman}  G.Bergman, Phys. Rep. {\bf 107}, 1 (1984).

\bibitem{Rama} P. A. Lee and T. V. Ramakrishnan, Rev. Mod. Phys. {\bf 57}, 287 (1987).

\bibitem{Fukuyama} P. A. Lee, A. D. Stone, and H. Fukuyama,
Phys. Rev. B {\bf 55}, 1039 (1987).

\bibitem{Imry} Y. Imry, Europhys. Lett. {\bf 1}, 249 (1986).


\bibitem{BeenakkerReview} C. W. J. Beenakker, Rev. Mod. Phys. {\bf 69}, 731 (1997)

\bibitem{cavity} R. Bl\"{u}mel and L. Smilansky , Phys. Rev. Lett. {\bf 60}, 477 (1988).

\bibitem{NazarovOhm} Yu.~V.~Nazarov, in: {\em Quantum Dynamics of
Submicron Structures}, ed. by H.~A.~Cerdeira, B.~Kramer, and
G.~Sch\"on, Kluwer Academic Publishers, Dordrecht (1995), p.687. 


\bibitem{Efetov} K.B. Efetov, Adv. Phys.  {\bf 32}, 53 (1983).   


\bibitem{Zirnbauer} M.R.Zirnbauer, Nucl. Phys B  {\bf 265} 375 (1986).

\bibitem{Iida}
 S. Iida, H. A. Weidenmueller, and J. A. Zuk, Ann. Phys. (N.Y.) {\bf 200}, 219 (1990).  

\bibitem{Carlo-corrections}
C. W. J. Beenakker, Phys. Rev. B 49, 2205 (1994);
C. W. J. Beenakker, Phys. Rev. Lett. 70, 4126 (1993).
\bibitem{Yuli-corrections}
Yu. V. Nazarov, Phys. Rev. B {\bf 52}, 4720–4723 (1995);
Yu. V. Nazarov, Phys. Rev. Lett. {\bf 76}, 2129 (1996). 

\bibitem{circuit-theory-old}
Yu. V. Nazarov, Phys. Rev. Lett. 73, 1420 (1994).  


\bibitem{fcs-yuli} Yu. V. Nazarov and D. A. Bagrets,  
Phys. Rev. Lett. {\bf 88}, 196801 (2002).

\bibitem{ct-yuli} Yu. V. Nazarov, Phys. Rev. Lett. {\bf 73}, 134 (1994).

\bibitem{spin}  A. Brataas, Yu. V. Nazarov, and G. E. W. Bauer 
Phys. Rev. Lett. 84, 2481 (2000) 

\bibitem{non-eq}Yu.~V. Nazarov, 
  Superlattices and Microst.\ {\bf 25}, 1221 (1999);
  cond-mat/9811155.

\bibitem{AGD} A. A. Abrikosov, L.P. Gor'kov, and I.E. Dzhyaloshinskii, Methods of Quantum Field Theory in Statistical Mechanics (Dover Publications) (1977).
\bibitem{Usadel} 
K.D.\ Usadel,
Phys.\ Rev.\ Lett.\ {\bf 25}, 507 (1970).
\bibitem{Larkin}
A. I. Larkin and Yu. V. Ovchinninkov, 
Sov. Phys. JETP {\bf 41},960 (1975);A. I. Larkin and Yu. V. Ovchinnikov, Sov. Phys. JETP
{\bf 46}, 155 (1977).
\bibitem{Kuprianov}
A. A. Golubov, M. Yu. Kupriyanov, and E. Il'ichev 
Rev. Mod. Phys. 76, 411 (2004). 
\bibitem{Luttinger}
J. M. Luttinger and J.C. Ward,  
Phys. Rev. {\bf 118}, 1417 (1960).
\bibitem{Larkin-great} M. V. Feigel'man, A. I. Larkin and M.
A. Skvortsov, Phys. Rev. B {\bf 61}, 12361 (2000).
\bibitem{Yuli-Annalen} Yu.~V. Nazarov, 
 Ann. Phys. (Leipzig) {\bf 8}, SI-193 (1999),
 cond-mat/9908143.
\bibitem{parametric} B. L. Altshuler and B. I. Shklovskii,
Zh. Exp. Theor. Phys. {\bf 91}, 220 (1986) [ Sov. Phys. JETP {\bf 64}, 127 (1986)].
\bibitem{remark} Although this fact is in principle known 
form \cite{Larkin} and therefore preceeds the discovery 
of $G_Q$ corrections, it is still not commonly recognized. 
On many occasions, the presence of Copperon ladders and
Hikami boxes in a diagrammatic theory has been confusingly
misinterpreted as "a signature of quatum corrections". 

\bibitem{altrev2}  B.~L.~Altshuler and  A.G. Aronov, in: Electron-Electron Interactions in Disordered Systems p.1, eds., A.L. Efros and M.Pollak (North Holland, Amsterdam, 1985).

\bibitem{ABGreview} I. L. Aleiner, P. Brouwer and L. I. Glazman, Physics Reports {\bf 358}, 309 (2002).

\bibitem{aron} A.G. Aronov and Y.V. Sharvin, Rev. Mod. Phys. {\bf 59}, 755 (1987). 

\bibitem{stoof} T. H. Stoof and Yu. V. Nazarov,  Phys. Rev. B {\bf 54}, R772 (1996).


\bibitem{piet} P.W. Brouwer and C.W.J. Beenakker, J. Math. Phys. {\bf 37}, 4904 (1996). 

\bibitem{Beenakker-QPC} C. W. J. Beenakker and J. A. Melsen, Phys. Rev. B {\bf 50}, 2450 (1994).

\bibitem{campagnano} G. Campagnano, O. N. Jouravlev, Ya. M. Blanter, and Yu. V. Nazarov,
Phys. Rev. B {\bf 69}, 235319 (2004)

\end{thebibliography}
\end{document}